\documentclass[prd,twocolumn,showpacs,amsmath,amssymb,nofootinbib,superscriptaddress]{revtex4}
\usepackage{graphicx}
\usepackage[pdftex]{color}
\usepackage{bm}
\usepackage{amsfonts}
\usepackage{amssymb}
\usepackage{color}
\usepackage{float}
\usepackage{amsmath}
\usepackage{dcolumn}
\usepackage{hyperref}
\usepackage{amsfonts}

\renewcommand{\[}{\left[}
\renewcommand{\]}{\right]}

\newcommand{\nuu}{\ensuremath{{\tilde{\nu}}}}

\def\be{\begin{equation}}
\def\ee{\end{equation}}
\def\bea{\begin{eqnarray}}
\def\eea{\end{eqnarray}}
\def\eqi{\begin{equation}}
\def\eqf{\end{equation}}
\def\eqia{\begin{eqnarray}}
\def\eqfa{\end{eqnarray}}

\newcommand{\fs}{{\rm{\it f\sigma_8}}}
\begin{document}

\title{Conjoined constraints on modified gravity from the expansion history and cosmic growth}

\author{Spyros Basilakos}\email{svasil@academyofathens.gr}
\affiliation{Academy of Athens, Research Center for Astronomy and Applied Mathematics, Soranou Efesiou 4, 11527, Athens, Greece}

\author{Savvas Nesseris}\email{savvas.nesseris@csic.es}
\affiliation{Instituto de F\'isica Te\'orica UAM-CSIC, Universidad Auton\'oma de Madrid, Cantoblanco, 28049 Madrid, Spain}

\date{\today}

\begin{abstract}
In this paper we present conjoined constraints on several cosmological models from the expansion history $H(z)$ and cosmic growth $f\sigma_8(z)$. The models we study include the CPL $w_0w_a$ parametrization, the Holographic Dark Energy (HDE) model, the Time varying vacuum ($\Lambda_t$CDM) model, the Dvali, Gabadadze and Porrati (DGP) and Finsler-Randers (FRDE) models, a power law $f(T)$ model and finally the  Hu-Sawicki $f(R)$ model. In all cases we perform a simultaneous fit to the SnIa, CMB, BAO, $H(z)$ and growth data, while also following the conjoined visualization of $H(z)$ and $f\sigma_8(z)$ as in Linder (2017). Furthermore, we introduce the Figure of Merit (FoM) in the $H(z)-f\sigma_8(z)$ parameter space as a way to constrain models that jointly fit both probes well. We use both the latest $H(z)$ and $f\sigma_8(z)$ data, but also LSST-like mocks with $1\%$ measurements and we find that the conjoined method of constraining the expansion history and cosmic growth simultaneously is able not only to place stringent constraints on these parameters but also to provide an easy visual way to discriminate cosmological models. Finally, we confirm the existence of a tension between the growth rate and Planck CMB data and we find that the FoM in the conjoined parameter space of $H(z)-f\sigma_8(z)$ can be used to discriminate between the $\Lambda$CDM model and certain classes of modified gravity models, namely the DGP and $f(T)$.
\end{abstract}

\maketitle

\section{Introduction}
The portrait of the cosmos, as it is revealed by the analysis of various independent cosmological observations (see Ref.\cite{Ade:2015xua} and references therein), is tightly tied with a spatially flat universe in which the cosmic fluid contains $\sim 30\%$ of matter (baryonic and dark) and the rest is the so called dark energy (DE). Although there is mounting observational evidence that DE is responsible for the accelerated expansion of the universe, the underlying mechanism behind such a phenomenon is yet unknown.

Despite the lack of our knowledge regarding the nature of the DE, in the literature there is a large class of cosmological models which mathematically treats the accelerated expansion of the Universe. In general these cosmological scenarios are split into two large groups. The first category of DE models adhere to General Relativity (GR) and propose the existence of new fields in nature (for review see \cite{Copeland:2006wr,Caldwell:2009ix} and references therein). The second group of cosmological models is mainly based on modified gravity for which the present accelerating era appears as a geometric effect due to the fact that gravity becomes weak at cosmological scales \cite{Copeland:2006wr,Clifton:2011jh}.
It is interesting to mention that in the context of modified gravity models the effective equation-of-state (EoS) parameter can enter in the phantom regime, namely $w<-1$.

At the perturbation level, the growth of matter fluctuation provides a useful tool to investigate the matter distribution in the Universe \cite{Okumura:2015lvp}, and, more importantly, it can be measured from observations. Indeed the growth rate data are mainly based on galaxy surveys, like SDSS, BOSS, {\em WiggleZ} etc (see our Table \ref{tab:fsigma8data} and references therein). The growth rate data have been used extensively in the literature in order to put constraints on the growth index $\gamma$. The measurement of the growth index provides an efficient way to discriminate between modified gravity models and DE models which are developed in the context of GR. Indeed, one can find a large family of studies in which the predicted growth index is given analytically for various DE models, including scalar field DE \cite{Silveira:1994yq,Wang:1998gt,Linder:2003dr,Linder:2004ng,Nesseris:2007pa}, DGP \cite{Lue:2004rj, Linder:2004ng,Gong:2008fh}, $f(R)$ \cite{Gannouji:2008wt,Tsujikawa:2009ku, Boubekeur:2014uaa}, Finsler-Randers \cite{Basilakos:2013ij}, time varying vacuum models $\Lambda(H)$, \cite{Basilakos:2015vra}), Clustered DE \cite{Mehrabi:2015hva}, Holographic dark energy \cite{Mehrabi:2015kta} and $f(T)$ \cite{Basilakos:2016xob}.

From the aforementioned discussion it becomes clear that up to now the expansion data and the growth data have been used separately in order to study the cosmic history at the background and perturbations levels respectively. In this article we follow a different path, namely the {\it conjoined method} recently proposed by Linder \cite{Linder:2016xer} where it was claimed that in order to distinguish the DE models it is better to use the Hubble parameter $H(z)$ directly with the growth quantity $f\sigma_{8}(z)$ in a conjoined diagram, rather than using each as a function of redshift. In the current article we attempt to investigate such a possibility by computing the $H-f\sigma_{8}$ diagram of the most popular DE models and comparing the corresponding predictions with the observed conjoined diagram provided by the cosmic chronometer $H(z)$ and the growth data. A similar, but not overlapping analysis, of the conjoined constraints was also made by Ref.~\cite{Moresco:2017hwt}, where the authors studied extensions of the $\Lambda$CDM model, such as a constant but free equation of state $w$ or the effect of neutrinos. However, we consider a much bigger ensemble of dark energy models, hence our analysis in much more broad in scope.

The layout of the article is as follows:
In Sec. II we discuss the linear growth of matter fluctuations in the dark energy regime, while in Sec. III we present the basic properties of various cosmological models, including those of modified gravity. In Sec. IV we discuss the tension between the growth and Planck15 CMB data, while in Sec. V, we test the DE cosmologies by comparing the theoretical predictions of the conjoined $H-\fs$ diagram with observations. Finally, we present our conclusions in Sec. VI.

\section{Linear growth of matter Perturbations}
In this section we present the main ingredients of the linear growth of matter perturbations within the context of different types of dark energy. Since we are in the matter dominated epoch we can neglect the radiation component from the cosmic expansion. Let us start the current analysis with the basic differential equation at subhorizon scales \cite{Lue:2004rj,Linder:2004ng,Gannouji:2008wt,Stabenau:2006td,Uzan:2006mf,Tsujikawa:2007tg}
\begin{equation}
\label{eq:111}
\ddot{\delta}_{m}+2\nuu H\dot{\delta}_{m}-4\pi G\mu \rho_{m} \delta_{m}=0 \;.
\end{equation}
As is well known, a general solution of the aforementioned equation is written as $\delta_{m} \propto D(t)$. Notice, that $D(t)$ is the growth factor usually scaled to unity at the present time. In order to address the issue of how the quantities $\nuu$ and $\mu\equiv G_{\rm eff}/G_{N}$ affect the matter fluctuations, one has to deal in general with the following three distinct scenarios:

\begin{enumerate}
  \item The situation in which the dark energy models (quintessence and the like) adhere to GR, hence $(\nuu,\mu)=(1,1)$.
  \item The case where $(\nuu,\mu)\ne (1,1)$ which implies that there are interactions in the dark sector and, finally.
  \item The case of either modified gravity models or inhomogeneous dark energy models (inside GR), hence
$\nuu=1$ and $\mu\ne1$\footnote{Note that if matter and geometry are coupled then we have $\mu\equiv G_{\rm eff}/G_{N}+\beta(a,k)$, in which $\beta(a,k)$ is a quantity that depends on derivatives of the Lagrangian of the model \cite{Nesseris:2008mq}.}.
\end{enumerate}

Now we are ready to introduce the growth rate of clustering (first proposed by Ref.~\cite{Peebles:1994xt}) as follows
\begin{equation}
\label{fzz221}
f(a)=\frac{d\ln \delta_{m}}{d\ln a}\simeq \Omega^{\gamma}_{m}(a)
\end{equation}
with
\begin{equation}
\label{ddomm}
\Omega_{m}(a)=\frac{\Omega_{m0}a^{-3}}{E^{2}(a)}
\end{equation}
where $E(a)=H(a)/H_{0}$ is the normalized Hubble parameter and $\gamma$ is the growth index. By differentiating Eq.(\ref{ddomm}) it is easy to show
\begin{equation}
\label{ddomm1}
\frac{d\Omega_{m}}{da}=-3\frac{\Omega_{m}(a)}{a}\left( 1+\frac{2}{3}
\frac{d{\rm ln}E}{d{\rm ln}a} \right) \;.
\end{equation}
Furthermore, combining Eqs.(\ref{fzz221})-(\ref{ddomm1}) the main equation (\ref{eq:111} becomes
\be
a\frac{df}{da}+ \left(2\nuu+\frac{d{\ln}E}{d{\rm ln}a}\right)f+f^{2}
=\frac{3\mu \Omega_{m}}{2},\label{fzz444}
\ee
or
{\small{
\begin{equation}
\label{Poll}
a{\rm ln}(\Omega_{m})\frac{d\gamma}{da}+\Omega_{m}^{\gamma}
-3\gamma+2\nuu-
\left(\gamma-\frac{1}{2}\right)\frac{d{\ln}E}{d{\rm ln}a}=\frac{3}{2}
\mu\Omega_ { m } ^ { 1-\gamma}.
\end{equation}}}
It is interesting to mention that Steigerwald et al. \cite{Steigerwald:2014ava} have provided
another expression of the above equation, namely
\be
\label{Ster}
\frac{d\omega}{d{\ln}a}(\gamma+\omega\frac{d\gamma}{d\omega}) +{\rm
e}^{\omega \gamma}+2\nuu+\frac{d{\ln}E}{d{\rm
ln}a}=\frac{3}{2}\mu {\rm e}^{\omega(1-\gamma)},
\ee
where $\omega={\rm ln}\Omega_{m}(a)$ which implies that at $z\gg 1$ ($a \to 0$) we get
$\Omega_{m}(a)\to 1$ [or $\omega \to 0$]. Based on Eq.(\ref{Ster})
Steigerwald et al. \cite{Steigerwald:2014ava} proposed a general mathematical approach
in order to derive the asymptotic value of the growth index which is given by
(see Eq.(8) in Ref.~\cite{Steigerwald:2014ava} and the relevant discussion in Ref.~\cite{Basilakos:2015vra})
\be
\label{g000}
\gamma_{\infty}=\frac{3(M_{0}+M_{1})-2(H_{1}+N_{1})}{2+2X_{1}+3M_{0}},
\ee
where %the relevant quantities are
\be \label{Coef1}
M_{0}=\left. \mu \right|_{\omega=0}\,,
\ \
M_{1}=\left.\frac{d \mu}{d\omega}\right|_{\omega=0}
\ee
and
\be \label{Coef2}
N_{1}=\left.\frac{d \nuu}{d\omega}\right|_{\omega=0}\,,\ \
H_{1}=-\frac{X_{1}}{2}=\left.\frac{d \left(d{\rm ln}E/d{\rm ln}a\right)}{d\omega}\right|_{\omega=0} \,.
\ee

Since the exact functional form of the growth index has yet to be found, here we utilize the well known Taylor expansion around $a(z)=1$ (see Ref.~\cite{Polarski:2007rr,Belloso:2011ms,DiPorto:2011jr})
\be
\label{aPoll1}
\gamma(a)=\gamma_{0}+\gamma_{1}(1-a),%=\gamma_{0}+\gamma_{1}\frac{z}{1+z} \;,
\ee
in which the asymptotic value boils down to $\gamma_{\infty}\simeq \gamma_{0}+\gamma_{1}$, where we have set
$\gamma_{0}=\gamma(1)$.
Lastly, writing Eq.(\ref{Poll}) at the present time ($a=1$)
\begin{eqnarray}
\label{Poll1}
\hspace{-0.4cm}&&-\gamma^{\prime}(1){\rm
ln}(\Omega_{m0})+\Omega_{m0}^{\gamma(1)}-3\gamma(1)+2\nuu_{0}
-2(\gamma_{0}-\frac{1}{2})\left.\frac{d{\ln}E}{d{\rm ln}a}\right|_{a=1}\nonumber\\
\hspace{-0.4cm}&& =\frac {3}{2} \mu_ { 0 } \Omega_{m0}^{1-\gamma(1)}, \ \ \
\end{eqnarray}
and with the aid of Eq.(\ref{aPoll1}) we arrive at
\begin{equation}
\label{Poll2}
\gamma_{1}=\frac{\Omega_{m0}^{\gamma_{0}}-3\gamma_{0}+2\nuu_{0}
-2(\gamma_{0}-\frac{1}{2})\left.\frac{d{\ln}E}{d{\rm ln}a}\right|_{a=1}
-\frac{3}{2}\mu_{0}\Omega_{m0}^{1-\gamma_{0}} }
{\ln  \Omega_{m0}}\;,
\end{equation}
where prime denotes a derivative with respect to the scale factor,
$\mu_{0}=\mu(1)$ and $\nuu_{0}=\nuu(1)$.

\section{Dark Energy Models}\label{sec.constr}
In this section we describe the ten distinct cosmological models used in our analysis. Notice that the observational viability of these DE models was recently tested in Basilakos and Nesseris \cite{Basilakos:2016nyg} using the JLA supernovae, the BAO data and the CMB shift parameters, but also the $H(z)$ data shown in Table \ref{tab:Hzdata}. We also used a growth-rate data compilation which we now update to the ``Gold-2017" set of Ref.~\cite{Nesseris:2017vor}, shown for completeness in Table \ref{tab:fsigma8data}. For the details of the analysis of the JLA, BAO, CMB and $H(z)$ data we refer the interested reader to Ref.~\cite{Basilakos:2016nyg}, while for that of the new growth-rate data we refer to Ref.~\cite{Nesseris:2017vor}. The best-fit parameters and their errors for all of the models used in this analysis were obtained via an MCMC, and the results are shown in Table~\ref{tab:bestfit}.\footnote{The codes used in the analysis are freely available at \url{www.uam.es/savvas.nesseris/}. }

Knowing the basic cosmological functions of a given dark energy model and the corresponding model parameters, it is trivial to compute the asymptotic value of the growth index $\gamma_{\infty}$ from Eq.~(\ref{g000}). The next step is to solve the system of $\gamma_{\infty}=\gamma_{0}+ \gamma_{1}$ and Eq.(\ref{Poll2}) in order
to compute $(\gamma_{0},\gamma_{1})$ from the cosmological parameters. Let us now briefly present the cosmological models explored in the current work. Notice, that in all cases we assume a spatially flat Friedmann-Lema\^\i tre-Robertson-Walker (FLRW) geometry.

\begin{itemize}

\item {\bf $w$CDM model.} In this case we consider as constant the equation of state (hereafter EoS) parameter $w=p_{d}/\rho_{d}$, where $p_{d}$ and $\rho_{d}$ are the pressure and density of the dark energy fluid respectively. Since this model is inside GR and does not allow interactions in the dark sector we have $\mu(a)=\nuu(a)=1$. The latter conditions imply that the dimensionless Hubble parameter is written as
\be
\label{EWCDM}
E^{2}(a)=\Omega_{m0}a^{-3}+\Omega_{d0}a^{-3(1+w)},
\ee
where the model parameters are $\Omega_{m0}=1-\Omega_{d0}$ and $w$. Using Eq.(\ref{EWCDM}) we find
\be
\label{EWCDM11}
\frac{d{\ln}E}{d{\rm ln}a}=
-\frac{3}{2}-\frac{3}{2}w\left[1-\Omega_{m}(a)\right]
\ee
and
$$
\{ M_{0},M_{1},H_{1},X_{1}\}=\{ 1,0,\frac{3w}{2},-3w \}\;.
$$
Inserting the above coefficients into Eq.(\ref{g000}) we recover the well-known asymptotic value of the growth index
(see also Ref.~\cite{Silveira:1994yq,Wang:1998gt,Linder:2003dr,Lue:2004rj,Linder:2004ng,Nesseris:2007pa,Basilakos:2012uu}), namely
$$
\gamma_{\infty} = \frac{3(w-1)}{6w-5}.
$$
As expected for $w=-1$ the $w$CDM model reduces to a $\Lambda$CDM cosmological model in which $\gamma_{\infty}^{(\Lambda)} = 6/11$.

\item {\bf CPL model ($w_{0}w_{a}$CDM).} This phenomenological model was first proposed by Chevalier-Polarski-Linder \cite{Chevallier:2000qy,Linder:2002et}. Specifically, the dark energy EoS parameter is given as a first order Taylor expansion around the present epoch, $w(a)=w_{0}+w_{1}(1-a)$. This model shares the same cosmological properties with $w$CDM which means that $\mu(a)=\nuu(a)=1$. The normalized Hubble parameter is given by
$$
E^{2}(a)=\Omega_{m0}a^{-3}+\Omega_{d0} a^{-3(1+w_{0}+w_{1})}e^{3w_{1}(a-1)}.
$$
where the free parameters of the models are $(\Omega_{m0},w_{0},w_{1})$ with $\Omega_{d0}=1-\Omega_{m0}$.
The quantity $d{\rm ln}E/d{\rm ln}a$ is given by Eq.(\ref{EWCDM11}) but here the EoS parameter is $w=w(a)$.
Within this context, the growth coefficients (see also Ref.~\cite{Steigerwald:2014ava}) are written as
$$
\{ M_{0},M_{1},H_{1},X_{1}\}=\{ 1,0,\frac{3(w_{0}+w_{1})}{2},-3(w_{0}+w_{1}) \}
$$
and thus
$$
\gamma_{\infty} = \frac{3(w_{0}+w_{1}-1)}{6(w_{0}+w_{1})-5} .
$$

\item {\bf Running $\Lambda$ ($\Lambda_{t}$CDM model).} Here we allow $\Lambda$ to vary with redshift. Using the notations of Refs.~\cite{Shapiro:2000dz,EspanaBonet:2003vk} the evolution of the vacuum is written as $\Lambda(H)=\Lambda_0+ 3\nu\,(H^{2}-H_0^2)$, where $\Lambda_0\equiv\Lambda(H_0)=3\Omega_{\Lambda 0}H^{2}_{0}$. In this case the normalized Hubble parameter takes the form
\begin{equation}
\label{anorm11}
E^{2}(a)=
{\tilde \Omega}_{\Lambda 0}+{\tilde \Omega_{m0}}a^{-3(1-\nu)}\;,
\end{equation}
with
\be
\label{EE1}
\frac{d{\ln}E}{d{\rm ln}a}=
-\frac{3}{2}(1-\nu){\tilde \Omega}_{m}(a)\;,
\ee
where we have set
${\tilde \Omega}_{m}(a)=\frac{\tilde \Omega_{m0}a^{-3(1-\nu)}}{E^{2}(a)}$,
${\tilde \Omega_{m 0}}\equiv \frac{\Omega_{m0}}{1-\nu}$ and
${\tilde \Omega_{\Lambda 0}}\equiv \frac{1-\Omega_{m 0}-\nu}{1-\nu}$.
Evidently, the free parameters of the model are $({\tilde \Omega_{m0}},\nu)$.
The basic quantities $\nuu$ and $\mu$ (see Ref.~\cite{Basilakos:2015vra}) are given by
\be
\label{nn1}
\nuu=1+\frac32\,\nu
\ee
and
\be
\label{mm1}
\mu(a)=
1-\nu-\frac{4\nu}{{\tilde \Omega}_{m}(a)}+3\nu(1-\nu).
\ee
Based on the above we compute the growth coefficients \cite{Basilakos:2015vra}
$$
\{ M_{0},M_{1},H_{1},X_{1}\}= \{ 1-2\nu-3\nu^{2},
-\frac{3(1-\nu)}{2},3(1-\nu) \}
$$
from which we provide
$$
\gamma_{\infty}=
\frac{6+3\nu}{11-12\nu}\;.
$$

\item {\bf Holographic dark energy (HDE) model.}
Using the holographic \cite{Ng:2000fq,Horava:2000tb} principle in GR $\nuu(a)=1$ one can show that
$$
w(a)=-\frac{1}{3}-\frac{2\sqrt{\Omega_{\rm d}(a)}}{3s}
$$
and
$$
\frac{d{\rm ln}\Omega_{d}}{d{\rm ln}a}=-\frac{w(a)}{3}\left[ 1-\Omega_{d}(a)\right]
$$
and
$$
E^2(a)=\frac{\Omega_{m0}a^{-3}}{1-\Omega_{d}(a)}\;,
$$
where $\Omega_{d}(a)=1-\Omega_{m}(a)$. Here the cosmological parameters are $\Omega_{m0}$ and $s$.
Note, that the expression of $d{\rm ln}E/d{\rm ln}a$ is given by Eq.(\ref{EWCDM}).
We would like to stress that the aforementioned three equations produce a system whose solution
provides $w(a)$, $\Omega_{d}(a)$ and $E(a)$.

The intrinsic features of the HDE are characterized by the quantity $\mu(a)$ as follows \cite{Mehrabi:2015kta}
\begin{equation} \label{VV}
\mu(a)=\left\{ \begin{array}{cc} 1
\;\;
       &\mbox{homogeneous HDE}\\
  1+\frac{\Omega_{\rm d}(a)}{\Omega_{\rm m}(a)}\Delta_{\rm d}(a)(1+3c_{\rm eff}^2)
\;\;
       & \mbox{clustered HDE}
       \end{array}
        \right.
\end{equation}
where $c_{\rm eff}^{2}$ is the effective sound speed of the dark energy and $\Delta_{d}=\frac{1+w(a)}{1-3w(a)}$ \cite{Batista:2013oca,Mehrabi:2015kta}.
For the homogeneous HDE model (hereafter HHDE)
we have (see also Ref.~\cite{Mehrabi:2015kta})
$$
\{ M_{0},M_{1},H_{1},X_{1}\}=\{ 1,0,\frac{3w_{\infty}}{2},-3w_{\infty} \}\;.
$$
where $w_{\infty}\simeq -1/3$, while the asymptotic value of the growth index becomes $\gamma_{\infty} = \frac{4}{7}$. The likelihood analysis of Ref.~\cite{Basilakos:2016nyg} provides $\Omega_{m0}=0.311\pm 0.003$ and $s=0.654\pm 0.006$ from which we get $(\gamma_{0},\gamma_{1}) \simeq (0.558,0.013)$.

Moreover, for clustered HDE (hereafter CHDE) we find
$$
\{ M_{0},M_{1},H_{1},X_{1}\}=\{ 1,-\frac{(1+3c_{\rm eff}^2)}{3},\frac{3{\rm w}_{\infty}}{2},-3{\rm w}_{\infty}\}
$$
and hence,
$$
\gamma_{\infty}=\frac{3(1-c_{\rm eff}^2)}{7} \;.
$$
Notice that we impose $c_{\rm eff}^{2}=0$, which means that the sound horizon is small with respect to
the Hubble radius and thus DE perturbations grow in a similar fashion to matter perturbations \cite{Garriga:1999vw,ArmendarizPicon:1999rj}.

\item {\bf $f(T)$ gravity model ($f_T$CDM model).} Among the large body of $f(T)$ gravity models here we focus on the power-law scenario first introduced by Bengochea and Ferraro \cite{Bengochea:2008gz}, with $f(T)=\alpha (-T)^{b}$, where $\alpha=(6H_0^2)^{1-b}\frac{\Omega_{F0}}{2b-1}$. In this case the dimensionless Hubble parameter is
\begin{eqnarray}
\label{Mod1Ezz}
E^2(a,b)=\Omega_{m0}a^{-3}+\Omega_{d0} E^{2b}(a,b)
\;,
\end{eqnarray}
and
\be
\label{Mod1Eza}
\frac{d{\ln}E}{d{\rm ln}a}=
-\frac{3}{2}\frac{\Omega_{m}(a)}{[1-bE^{2(b-1)}\Omega_{d0}]}
\ee
where $\Omega_{d0}=1-\Omega_{m0}$. Since $b\ll 1$ \cite{Linder:2010py} we use the approximation of Nesseris et al. \cite{Nesseris:2013jea} in order to simplify Eqs.(\ref{Mod1Ezz}) and (\ref{Mod1Eza}), namely
\be
\label{approxM1}
E^2(a,b)\simeq E^2_\Lambda(a)+\Omega_{d0}\ln\left[E^2_\Lambda(a)\right]b+... \;,
\ee
\be
\label{Taylor2}
\frac{d{\ln}E}{d{\rm ln}a}\simeq
-\frac{3}{2}\Omega_{m}(a)\left[ 1+\frac{\Omega_{d0}b}{E^{2}_{\Lambda}(a)}+...\right].
\ee
For this geometrical model we have ${\tilde \nu}=1$, while the quantity
$\mu$ is given by (see Ref.~\cite{Basilakos:2016xob} and references therein)
\begin{equation}
\mu(a)=\frac{1}{1+ \frac{b\Omega_{d0}} {(1-2b)E^{2(1-b)}}}
\end{equation}
or
\be
\label{Geff1}
\mu(a)\simeq 1-\frac{\Omega_{d0}}{E^{2}_{\Lambda}(a)}\;b+ \cdots .
\ee
Using the above expressions we are ready to compute the growth index coefficients of
(\ref{Coef1}) and (\ref{Coef2}) [see also Ref.~\cite{Basilakos:2016xob}]
$$
\{ M_{0},M_{1},H_{1},X_{1}\}= \{ 1,b,-\frac{3(1-b)}{2},3(1-b)\}
$$
from which we derive
$$
\gamma_{\infty} = \frac{6}{11-6b} \;.
$$
Obviously, for $b=0$ we recover the $\Lambda$CDM value $6/11$.

\item {\bf $f(R)$ gravity ($f_R$CDM model).}
We would like to finish the presentation of the cosmological models with the $f(R)$ model introduced by Hu and Sawicki. As proposed by Basilakos et al. \cite{Basilakos:2013nfa}, the Lagrangian of the current $f(R)$ model can be equivalently written as
\be
\label{Hu1}
f(R)= R- \frac{2\Lambda }{1+\left(\frac{b \Lambda }{R}\right)^n}
\ee
where $n$ is a parameter of the model (we set it to unity without loss of generality). Following the methodology of Ref.~\cite{Basilakos:2013nfa} the Hubble parameter is given in terms of a series expansion of the solution
of the equations on motion around $b=0$, i.e. $\Lambda$CDM, as
\be
H^2(a)=H_{\Lambda}^2(a)+\sum_{i=1}^M b^i \delta H_i^2(a), \label{expansion1}
\ee
where $H_{\Lambda}$ is the Hubble parameter of the concordance $\Lambda$CDM model. Notice, that $\delta H_i^2(a)$ is a set of algebraic quantities that can be determined from the equations of motion, see Ref.~\cite{Basilakos:2013nfa}. Moreover, the latter authors found that if we keep the two first non-zero terms ($M=2$) of the series then the approximated Hubble parameter is in excellent agreement with that provided by the numerical solution. Investigating the growth index in this model is not an easy task, since the modified Newton's constant is a function of both the scale factor $a$ and the scale $k$, ie $G_{\rm eff}=G_{\rm eff}(a,k)$ \cite{Tsujikawa:2007gd}. In particular, we have
\be
\mu(a,k)=\frac{G_{\rm eff}(a,k)}{G_N}=\frac{1}{F}\frac{1+4\frac{k^2}{a^2}F_{,R}/F}{1+3\frac{k^2}{a^2}F_{,R}/F},\label{geff}
\ee
where $F=f'(R)$, $F_{,R}=f''(R)$ and we have scaled Eq.~(\ref{geff}) so that in the case of $b=0$ ($\Lambda$CDM) we recover $\frac{G_{\rm eff}(a,k)}{G_N}=1$ as we should. Here we follow Ref.~\cite{Basilakos:2013nfa} and set $k=0.1h {\rm Mpc}^{-1}\simeq 300 H_0$, while $\tilde{\nu}(a)=1$.

Finally, in Ref.~\cite{Gannouji:2008wt} one can see that the current $f(R)$ model predicts rather low and rather high values for the growth index parameters, $(\gamma_0,\gamma_1)\simeq(0.4,-0.2)$. Due to the $k$-dependence of the effective Newton's constant in order to obtain the exact values of $(\gamma_0,\gamma_1)$ we first need to solve Eq.~(\ref{fzz444}) numerically in order to compute $\gamma_0\simeq\frac{\ln(f(1))}{\ln(\Omega_{m0})}$, where $f(1)$ is the growth rate at $a=1$, and then one can utilize Eq.~(\ref{Poll2}) to estimate $\gamma_1$.

\item {\bf Dvali, Gabadadze and Porrati (DGP) gravity model.}
In the framework of modified gravity models we first introduce that of Dvali, Gabadadze and Porrati \cite{Dvali:2000hr}. It is well known that the normalized Hubble parameter takes the form
\be
\label{eos222g}
E(a)=\sqrt{\Omega_{m0} a^{-3}+\Omega_{rc}}+\sqrt{\Omega_{rc}},
\ee
and thus
$\Omega_{rc}=(1-\Omega_{m0})^{2}/4$.
\be
\label{eos222e}
\frac{d{\rm ln}E}{d{\rm ln}a}=
-\frac{3\Omega_{m}(a)}{1+\Omega_{m}(a)}.
\ee
Here the quantity $\mu(a)=G_{\rm eff}/G_{N}$ takes the form
$$
\mu(a)=\frac{2+4\Omega^{2}_{m}(a)}{3+3\Omega^{2}_{m}(a)}
$$
and $\nuu(a)=1$.
Combining the above equations with Eqs.(\ref{Coef1}), (\ref{Coef2}) and (\ref{g000})  we obtain
$$
\{ M_{0},M_{1},H_{1},X_{1}\}=\{ 1,\frac{1}{3},-\frac{3}{4},\frac{3}{2}\}
$$
and $\gamma_{\infty}=\frac{11}{16}$ (see also Refs.~\cite{Linder:2004ng,Gong:2008fh}).
Similar to the $\Lambda$CDM model, the DGP gravity model contains one free parameter, namely $\Omega$.

\item {\bf Finsler-Randers dark energy model (FRDE).} The Finsler-Randers version of Finsler geometry has
been used in order to build the FRDE model (see Ref.~\cite{Stavrinos:2006rf} and references therein). Interestingly, it has been shown that the cosmic expansion of this model coincides with that of DGP gravity \cite{Basilakos:2013ij}.  Therefore, Eqs.(\ref{eos222g}) and (\ref{eos222e}) are also valid here and thus we use $\Omega_{m}=0.392\pm 0.008$.
Although the two dark energy models (FRDE and DGP) are identical at the background level, they deviate at the perturbation level because for the FRDE model we are dealing with $\mu(a)={\tilde \nu}(a)=1$ \cite{Stavrinos:2006rf}.
In this case it has been found that \cite{Basilakos:2016nyg} that
$$
\{ M_{0},M_{1},H_{1},X_{1}\}=\{ 1,0,-\frac{3}{4},\frac{3}{2}\}
$$
which implies $\gamma_{\infty}= \frac{9}{16}$.
\end{itemize}

\section{Tension with Planck and previous analyses}
At this point, we should stress that by performing the joint analysis of all the data, including the updated growth compilation ``Gold-2017" of Ref.~\cite{Nesseris:2017vor}, we confirm the existence of a tension between the growth rate and the Planck15 data which have been already observed, and also for other low redshift probes, and those discussed in the literature (see Refs.~\cite{Joudaki:2016mvz,Nesseris:2017vor,Basilakos:2014yda}).

This tension was not so obvious in the previous analysis of Basilakos and Nesseris \cite{Basilakos:2016nyg} for two reasons: First, the analysis was done in two steps by initially fitting the background to the SnIa, CMB, BAO and $H(z)$ data and then fixing these best-fit parameters $(\Omega_m,\Omega_b h^2,\cdots)$ before fitting the growth rate data. Second, not only were the growth rate data used were an older compilation with much bigger errors but the corrections required due to the Alcock-Paczynski effect, due to the different fiducial cosmologies assumed in the derivations of the data, were not implemented at the time, unlike in the current analysis or that of Ref.~\cite{Nesseris:2017vor}. These corrections contribute only up to a few percent, but could in principle bias the results and are thus necessary.

\begin{figure}[!t]
\centering
\includegraphics[width = 0.52\textwidth]{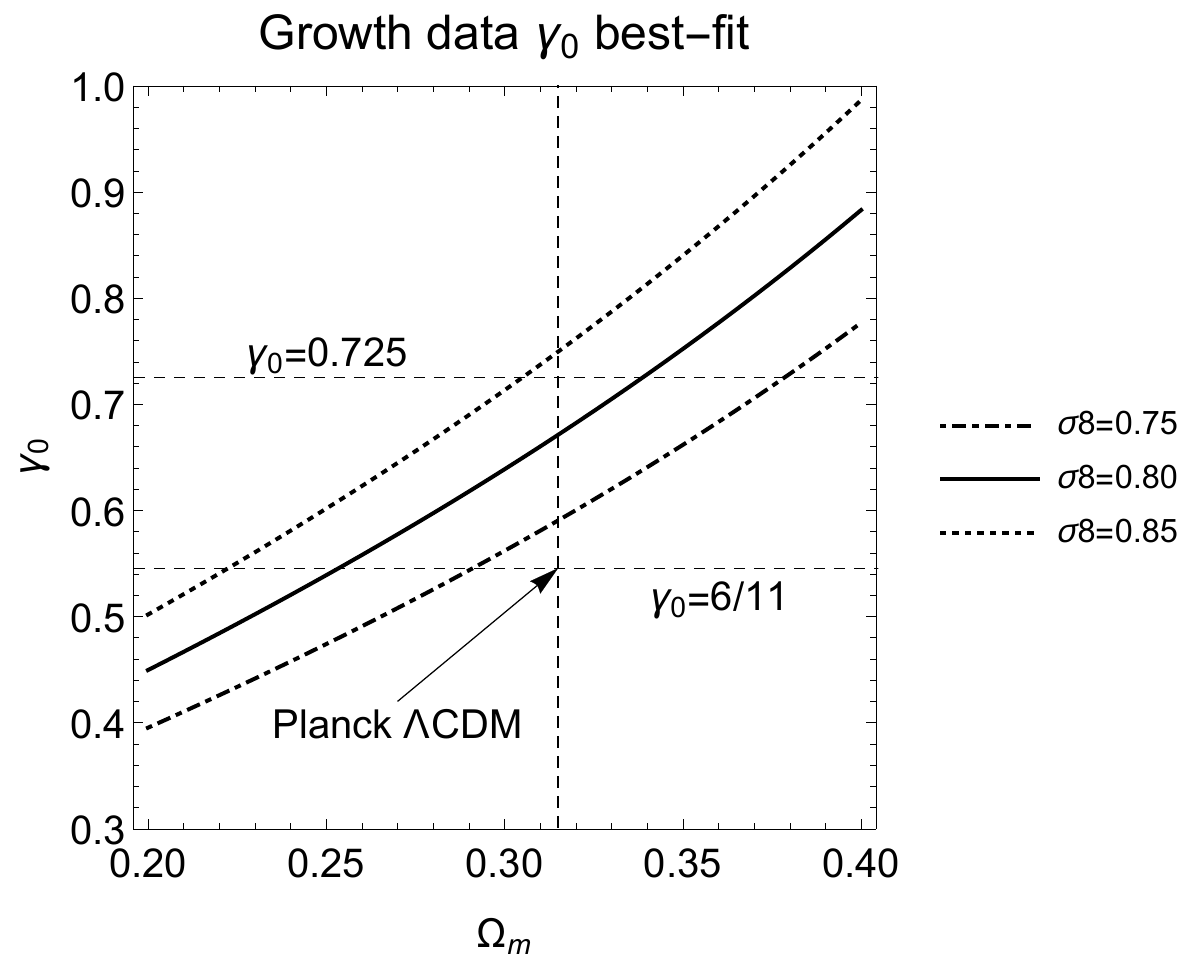}
\caption{The best-fit $\gamma_0$ for the $\Lambda$CDM model as a function of $\Omega_m$ for various values of the $\sigma_8$ parameter. As can be seen, the $f\sigma_8$ data push the best-fit $\gamma_0$ to higher values for higher $\Omega_m$. This means that when a Planck cosmology of $\Omega_m=0.315$ is used (shown with a vertical dashed line), then the best-fit is in the region of $\gamma_0\sim0.73$ in contrast to lower values of $\Omega_m$ that prefer a value for the $\gamma_0$ parameter closer to 6/11.}
\label{fig:plotg0}
\end{figure}

In order to understand why this tension but also the difference in the best-fit values of $(\gamma_0,\gamma_1)$ from Ref.~\cite{Basilakos:2016nyg} occurs, we show in Fig.~\ref{fig:plotg0} the best-fit $\gamma_0$ for the $\Lambda$CDM model as a function of $\Omega_m$ for various values of the $\sigma_8$ parameter. As can be seen, for higher values of $\Omega_m$ the new $f\sigma_8$ data push the best-fit $\gamma_0$ to higher values as well or in other words, they are positively correlated. This means that when the Planck cosmology of $\Omega_m=0.315$ is used (shown in Fig.~\ref{fig:plotg0} with a vertical dashed line), then the best-fit growth index is in the range of $\gamma_0\sim0.73$ in contrast to lower values of $\Omega_m$ that prefer a value for the $\gamma_0$ parameter closer to $6/11\simeq0.545$.

Furthermore, with the small arrow we show the Planck $\Lambda$CDM values of $(\gamma_0, \Omega_m)=(6/11, 0.315)$, which is quite far from our best-fit of $\gamma_0=0.725$. Also, it is important to note that the effect of the parameter $\sigma_8$ is quite strong, as it can strongly affect the value of the growth index. For example, as shown in Fig.~\ref{fig:plotg0}, by changing $\sigma$ between $\sigma_8\in\[0.75,0.85\]$ then the growth index changes by roughly $30\%$. This is also part of the reason for the disagreement with Ref.~\cite{Basilakos:2016nyg}, as the best-fit values of $\sigma_8$ in that case were quite lower than in this analysis.

Finally, this tension could be attributed to systematics in the low redshift probes \cite{Joudaki:2016mvz}, new physics in the form of modifications of gravity and $G_{eff}$ in particular \cite{Nesseris:2017vor}, or even in the form of a suppression of power at small scales \cite{Kunz:2015oqa}.

\begin{table}[!t]
\caption{The $H(z)$ data used in the current analysis (in units of $\textrm{km}~\textrm{s}^{-1} \textrm{Mpc}^{-1}$). This compilation is based partly in those of Refs.~\cite{Moresco:2016mzx} and \cite{Guo:2015gpa}.}
\label{tab:Hzdata}
\small
\centering
\begin{tabular}{cccccccccc}
\\
\hline\hline
$z$  & $H(z)$ & $\sigma_{H}$ & Ref.   \\
\hline
$0.07$    & $69.0$   & $19.6$  & \cite{Zhang:2012mp}  \\
$0.09$    & $69.0$   & $12.0$  & \cite{STERN:2009EP} \\
$0.12$    & $68.6$   & $26.2$  & \cite{Zhang:2012mp}  \\
$0.17$    & $83.0$   & $8.0$   & \cite{STERN:2009EP}    \\
$0.179$   & $75.0$   & $4.0$   & \cite{MORESCO:2012JH}   \\
$0.199$   & $75.0$   & $5.0$   & \cite{MORESCO:2012JH}   \\
$0.2$     & $72.9$   & $29.6$  & \cite{Zhang:2012mp}   \\
$0.27$    & $77.0$   & $14.0$  & \cite{STERN:2009EP}   \\
$0.28$    & $88.8$   & $36.6$  & \cite{Zhang:2012mp}  \\
$0.35$    & $82.7$   & $8.4$   & \cite{Chuang:2012qt}   \\
$0.352$   & $83.0$   & $14.0$  & \cite{MORESCO:2012JH}   \\
$0.3802$  & $83.0$   & $13.5$  & \cite{Moresco:2016mzx}   \\
$0.4$     & $95.0$   & $17.0$  & \cite{STERN:2009EP}    \\
$0.4004$  & $77.0$   & $10.2$  & \cite{Moresco:2016mzx}   \\
$0.4247$  & $87.1$   & $11.2$  & \cite{Moresco:2016mzx}   \\
$0.44$    & $82.6$   & $7.8$   & \cite{Blake:2012pj}   \\
$0.44497$ & $92.8$   & $12.9$  & \cite{Moresco:2016mzx}   \\
$0.4783$  & $80.9$   & $9.0$   & \cite{Moresco:2016mzx}   \\
$0.48$    & $97.0$   & $62.0$  & \cite{STERN:2009EP}   \\
$0.57$    & $96.8$   & $3.4$   & \cite{Anderson:2013zyy}   \\
$0.593$   & $104.0$  & $13.0$  & \cite{MORESCO:2012JH}  \\
$0.60$    & $87.9$   & $6.1$   & \cite{Blake:2012pj}   \\
$0.68$    & $92.0$   & $8.0$   & \cite{MORESCO:2012JH}    \\
$0.73$    & $97.3$   & $7.0$   & \cite{Blake:2012pj}   \\
$0.781$   & $105.0$  & $12.0$  & \cite{MORESCO:2012JH} \\
$0.875$   & $125.0$  & $17.0$  & \cite{MORESCO:2012JH} \\
$0.88$    & $90.0$   & $40.0$  & \cite{STERN:2009EP}   \\
$0.9$     & $117.0$  & $23.0$  & \cite{STERN:2009EP}   \\
$1.037$   & $154.0$  & $20.0$  & \cite{MORESCO:2012JH} \\
$1.3$     & $168.0$  & $17.0$  & \cite{STERN:2009EP}   \\
$1.363$   & $160.0$  & $33.6$  & \cite{Moresco:2015cya}  \\
$1.43$    & $177.0$  & $18.0$  & \cite{STERN:2009EP}   \\
$1.53$    & $140.0$  & $14.0$  & \cite{STERN:2009EP}  \\
$1.75$    & $202.0$  & $40.0$  & \cite{STERN:2009EP}  \\
$1.965$   & $186.5$  & $50.4$  & \cite{Moresco:2015cya}  \\
$2.34$    & $222.0$  & $7.0$   & \cite{Delubac:2014aqe}   \\
\hline\hline
\end{tabular}
\end{table}

\begin{table*}[t!]
\caption{A compilation of robust and independent $f\sigma_8(z)$ measurements from different surveys, compiled in Ref.~\cite{Nesseris:2017vor}. In the columns we show in ascending order with respect to redshift, the name and year of the survey that made the measurement, the redshift and value of $f\sigma_8(z)$ and the corresponding reference and fiducial cosmology. These datapoints are used in our analysis in the next sections.
\label{tab:fsigma8data}}
\begin{centering}
\begin{tabular}{ccccccc}
Index & Dataset & $z$ & $f\sigma_8(z)$ & Refs. & Year & Notes \\
\hline
1 & 6dFGS+SnIa & $0.02$ & $0.428\pm 0.0465$ & \cite{Huterer:2016uyq} & 2016 & $(\Omega_m,h,\sigma_8)=(0.3,0.683,0.8)$ \\

2 & SnIa+IRAS &0.02& $0.398 \pm 0.065$ &  \cite{Turnbull:2011ty},\cite{Hudson:2012gt} & 2011& $(\Omega_m,\Omega_K)=(0.3,0)$\\

3 & 2MASS &0.02& $0.314 \pm 0.048$ &  \cite{Davis:2010sw},\cite{Hudson:2012gt} & 2010& $(\Omega_m,\Omega_K)=(0.266,0)$ \\

4 & SDSS-veloc & $0.10$ & $0.370\pm 0.130$ & \cite{Feix:2015dla}  &2015 &$(\Omega_m,\Omega_K)=(0.3,0)$ \\

5 & SDSS-MGS & $0.15$ & $0.490\pm0.145$ & \cite{Howlett:2014opa} & 2014& $(\Omega_m,h,\sigma_8)=(0.31,0.67,0.83)$ \\

6 & 2dFGRS & $0.17$ & $0.510\pm 0.060$ & \cite{Song:2008qt}  & 2009& $(\Omega_m,\Omega_K)=(0.3,0)$ \\

7 & GAMA & $0.18$ & $0.360\pm 0.090$ & \cite{Blake:2013nif}  & 2013& $(\Omega_m,\Omega_K)=(0.27,0)$ \\

8 & GAMA & $0.38$ & $0.440\pm 0.060$ & \cite{Blake:2013nif}  & 2013& \\

9 &SDSS-LRG-200 & $0.25$ & $0.3512\pm 0.0583$ & \cite{Samushia:2011cs} & 2011& $(\Omega_m,\Omega_K)=(0.25,0)$  \\

10 &SDSS-LRG-200 & $0.37$ & $0.4602\pm 0.0378$ & \cite{Samushia:2011cs} & 2011& \\

11 &BOSS-LOWZ& $0.32$ & $0.384\pm 0.095$ & \cite{Sanchez:2013tga}  &2013 & $(\Omega_m,\Omega_K)=(0.274,0)$ \\

12 & SDSS-CMASS & $0.59$ & $0.488\pm 0.060$ & \cite{Chuang:2013wga} &2013& $\ \ (\Omega_m,h,\sigma_8)=(0.307115,0.6777,0.8288)$ \\

13 &WiggleZ & $0.44$ & $0.413\pm 0.080$ & \cite{Blake:2012pj} & 2012&$(\Omega_m,h)=(0.27,0.71)$ \\

14 &WiggleZ & $0.60$ & $0.390\pm 0.063$ & \cite{Blake:2012pj} & 2012& \\

15 &WiggleZ & $0.73$ & $0.437\pm 0.072$ & \cite{Blake:2012pj} & 2012 &\\

16 &Vipers PDR-2& $0.60$ & $0.550\pm 0.120$ & \cite{Pezzotta:2016gbo} & 2016& $(\Omega_m,\Omega_b)=(0.3,0.045)$ \\

17 &Vipers PDR-2& $0.86$ & $0.400\pm 0.110$ & \cite{Pezzotta:2016gbo} & 2016&\\

18 &FastSound& $1.40$ & $0.482\pm 0.116$ & \cite{Okumura:2015lvp}  & 2015& $(\Omega_m,\Omega_K)=(0.270,0)$\\
\end{tabular}\par\end{centering}
\end{table*}

\begin{table*}[t!]
\caption{The best-fit parameters of the various parameters of the models presented in the previous section.\label{tab:bestfit}}
\begin{centering}
{\scriptsize \hspace*{-1.5cm} \begin{tabular}{ccccccccccc}
Model & $\alpha$ & $\beta$ & $\Omega_m$ & $\Omega_b h^2$ & $\gamma_0$ & $\gamma_1$ & DE params & $h$ & $\sigma_8$ & $\chi^2_{min}$ \\
\hline
$\Lambda$CDM & $0.141\pm0.006$ & $3.102\pm0.004$ & $0.315\pm0.003$ & $0.0222\pm0.0001$ & $0.725\pm0.036$ & $0.584\pm0.127$ & ${}^{w_0=-1}_{w_a=0}$ & $0.673\pm0.002$ & $0.877\pm0.034$ & 743.303 \\
$w$CDM & $0.141\pm0.005$ & $3.097\pm0.005$ & $0.318\pm0.004$ & $0.0223\pm0.0004$ & $0.731\pm0.069$ & $0.581\pm0.058$ & ${}^{w_0=-0.987\pm0.005}_{w_a=0}$ & $0.670\pm0.003$ & $0.879\pm0.033$ & 743.235 \\
CPL & $0.141\pm0.004$ & $3.100\pm0.005$ & $0.317\pm0.006$ & $0.0222\pm0.0001$ & $0.729\pm0.036$ & $0.576\pm0.118$ & ${}^{w_0=-0.984\pm0.008}_{w_a=-0.020\pm0.004}$ & $0.671\pm0.004$ & $0.878\pm0.036$ &743.248 \\
DGP & $0.136\pm0.006$ & $3.076\pm0.017$ & $0.390\pm0.006$ & $0.0229\pm0.0001$ & $1.053\pm0.074$ & $0.900\pm0.092$ & $-$ & $0.588\pm0.003$ & $1.028\pm0.046$ &844.348\\
Time Var. & $0.141\pm0.008$ & $3.076\pm0.008$ & $0.319\pm0.004$ & $0.0222\pm0.0001$ & $0.737\pm0.064$ & $0.669\pm0.123$ & ${}^{s=(-2.95\pm13.6) \cdot 10 ^{-5}}$ & $0.671\pm0.004$ & $0.886\pm0.058$& 743.689\\
Holo & $0.142\pm0.005$ & $3.092\pm0.008$ & $0.314\pm0.003$ & $0.0224\pm0.0001$ & $0.740\pm0.055$ & $0.832\pm0.055$ & $v=0.669\pm0.008$ & $0.669\pm0.002$ & $0.896\pm0.047$ &751.237\\
$f(R)$ & $0.141\pm0.005$ & $3.110\pm0.013$ & $0.317\pm0.005$ & $0.0222\pm0.0001$ & $0.753\pm0.041$ & $0.690\pm0.062$ & $b=0.083\pm0.005$ & $0.671\pm0.003$ & $0.896\pm0.032$ & 743.337\\
$f(T)$ & $0.141\pm0.008$ & $3.095\pm0.027$ & $0.317\pm0.005$ & $0.0222\pm0.0001$ & $0.730\pm0.091$ & $0.586\pm0.067$ & $b=0.021\pm0.013$ & $0.671\pm0.003$ & $0.879\pm0.057$ &743.230\\
\end{tabular}}%\par%
\end{centering}
\end{table*}

\section{Conjoined Analysis with real and mock data}
In this section we implement the {\it conjoined method} of Ref.~\cite{Linder:2016xer} in order to distinguish the dark energy models, by placing constraints simultaneously on both the expansion history $H(z)$ and the growth of structure $f\sigma_{8}(z)$. Specifically, Ref.~\cite{Linder:2016xer} proposed to utilize the expansion history together with the cosmic growth history, namely the joined analysis of $H-f\sigma_{8}$, as a tool for testing the performance of the dark energy models. Obviously, in order to obtain the conjoined histories diagram it is essential to provide a proper collection of the $H-f\sigma_{8}$ data. Concerning the cosmic history we use the latest cosmic chronometer $H(z)$ data (see Table \ref{tab:Hzdata} and references therein), while for the growth data we refer the reader to our Table \ref{tab:fsigma8data}.

\begin{table}[!t]
%\centering
\caption{The binned $H(z)-f\sigma_8(z)$ growth data based on the original data of Tables \ref{tab:Hzdata} and \ref{tab:fsigma8data}.\label{tab:binned}}
\begin{tabular}{ccc}
%\hline
%\hline
Bins & $H(z)$ & $f\sigma_8(z)$ \\
\hline
$0 \leq z < 0.28$     \hspace{5pt}&\hspace{5pt} $75.489 \pm 2.718$ \hspace{5pt}&\hspace{5pt} $ 0.394\pm 0.023$ \\
$0.28 \leq z < 0.56$  \hspace{5pt}&\hspace{5pt} $83.572 \pm 3.501$ \hspace{5pt}&\hspace{5pt} $ 0.443\pm 0.028$ \\
$0.56 \leq z < 0.84$  \hspace{5pt}&\hspace{5pt} $95.540 \pm 2.482$ \hspace{5pt}&\hspace{5pt} $ 0.449\pm 0.036$ \\
$0.84 \leq z < 1.12$  \hspace{5pt}&\hspace{5pt} $129.189 \pm 10.862$ \hspace{5pt}&\hspace{5pt} $ 0.400\pm 0.110$ \\
%\hline
%\hline
\end{tabular}
\end{table}

\begin{figure*}[!t]
\centering
\includegraphics[width = 0.45\textwidth]{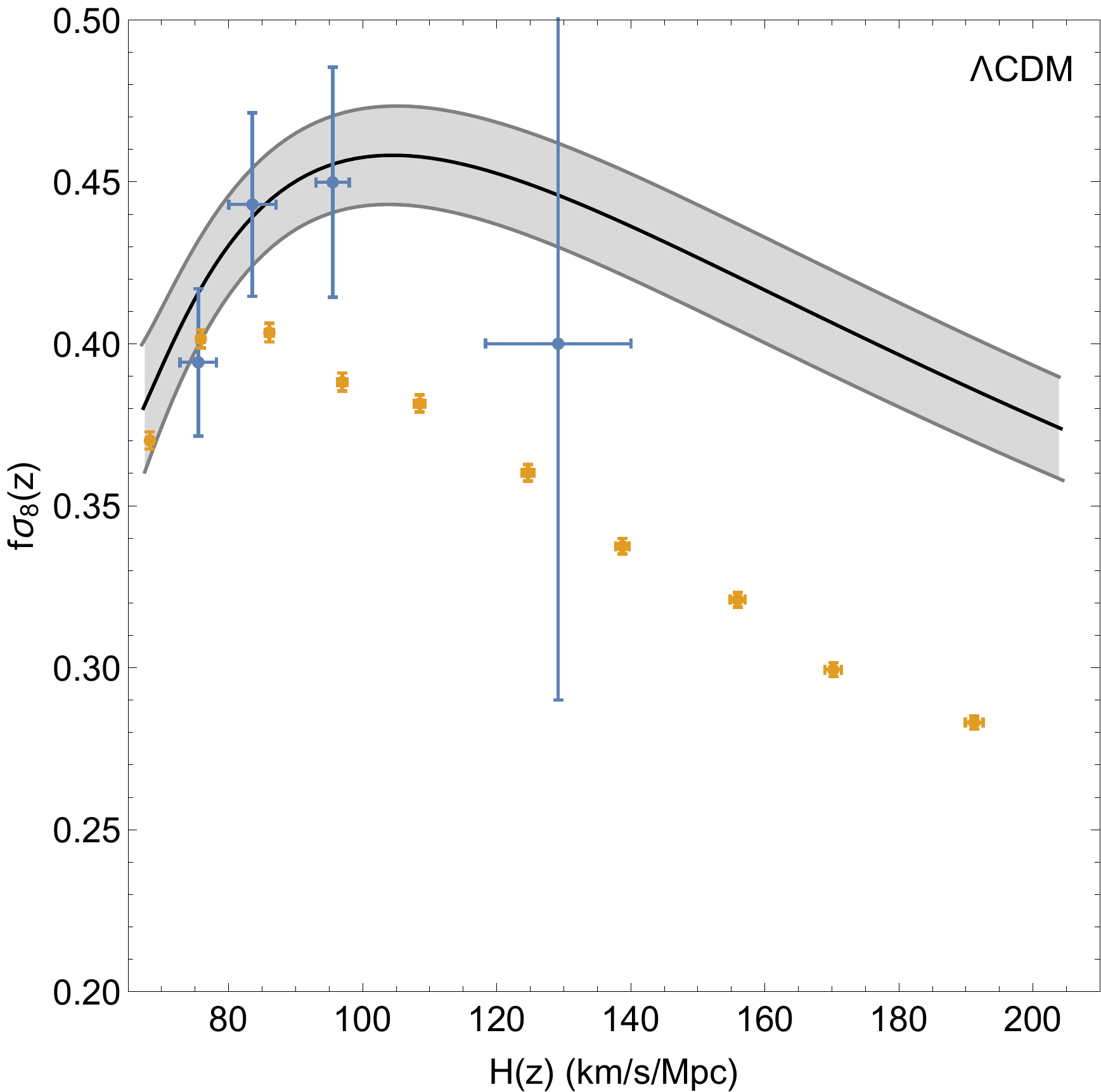}
\includegraphics[width = 0.45\textwidth]{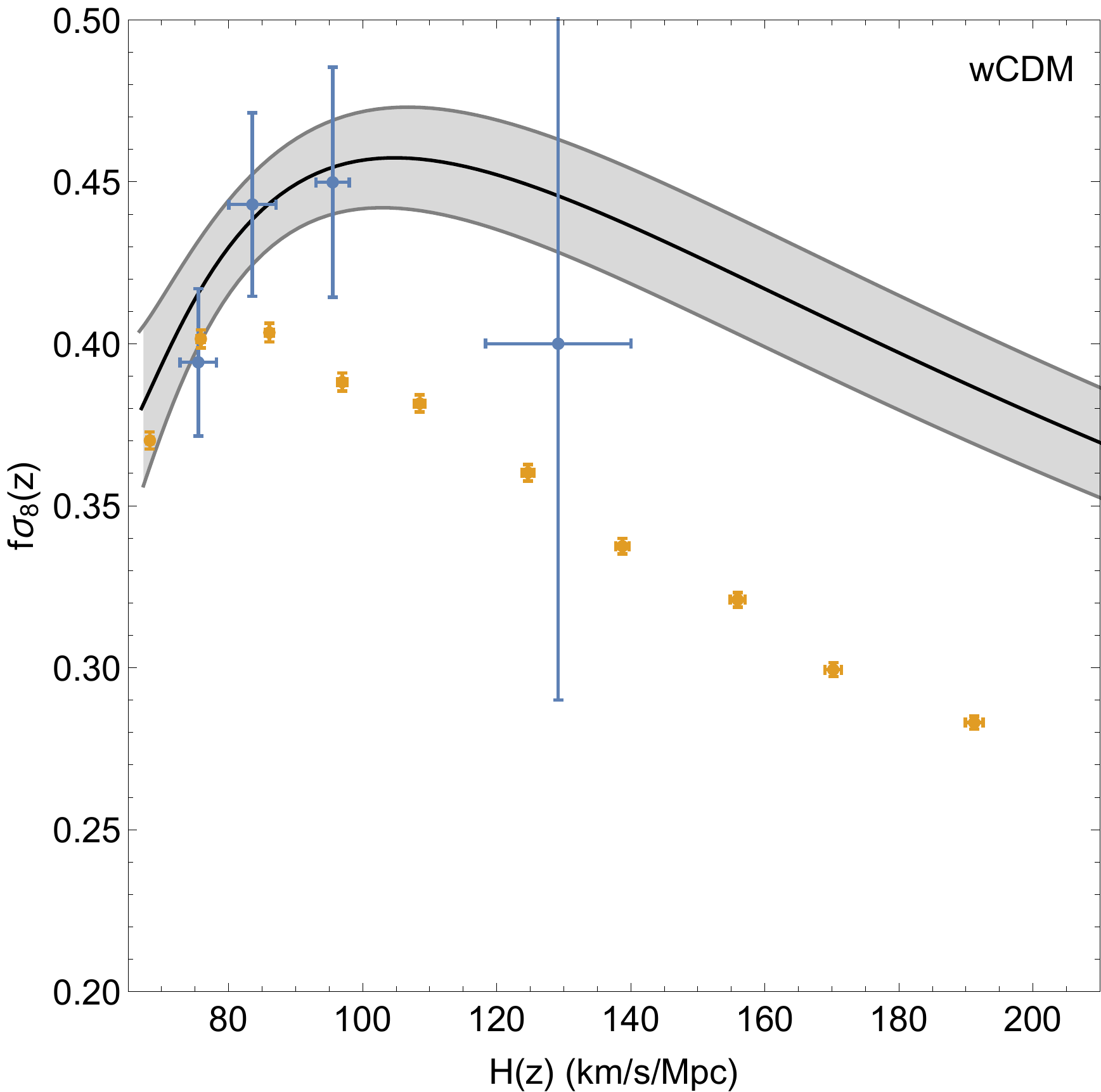}
\includegraphics[width = 0.45\textwidth]{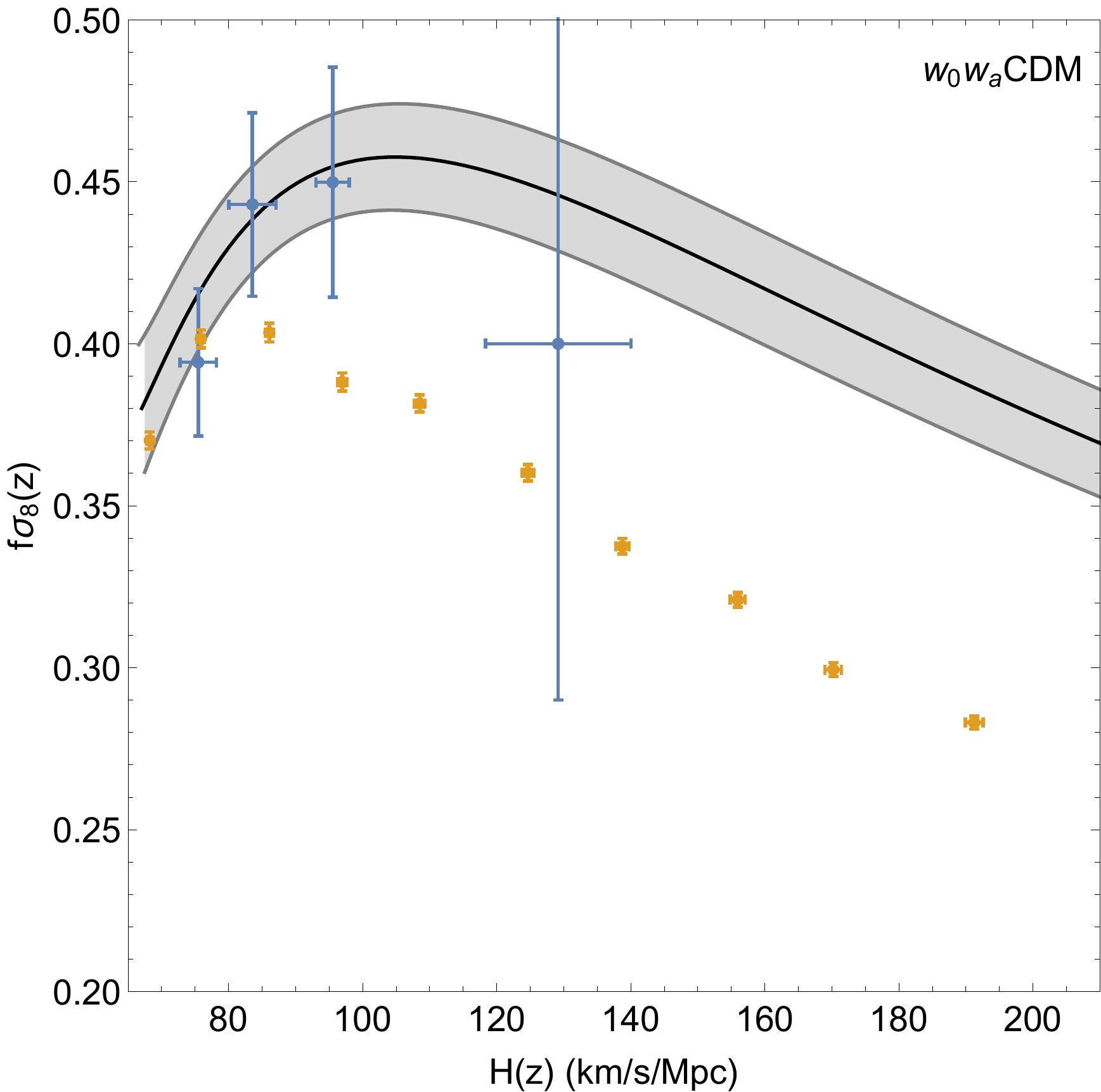}
\includegraphics[width = 0.45\textwidth]{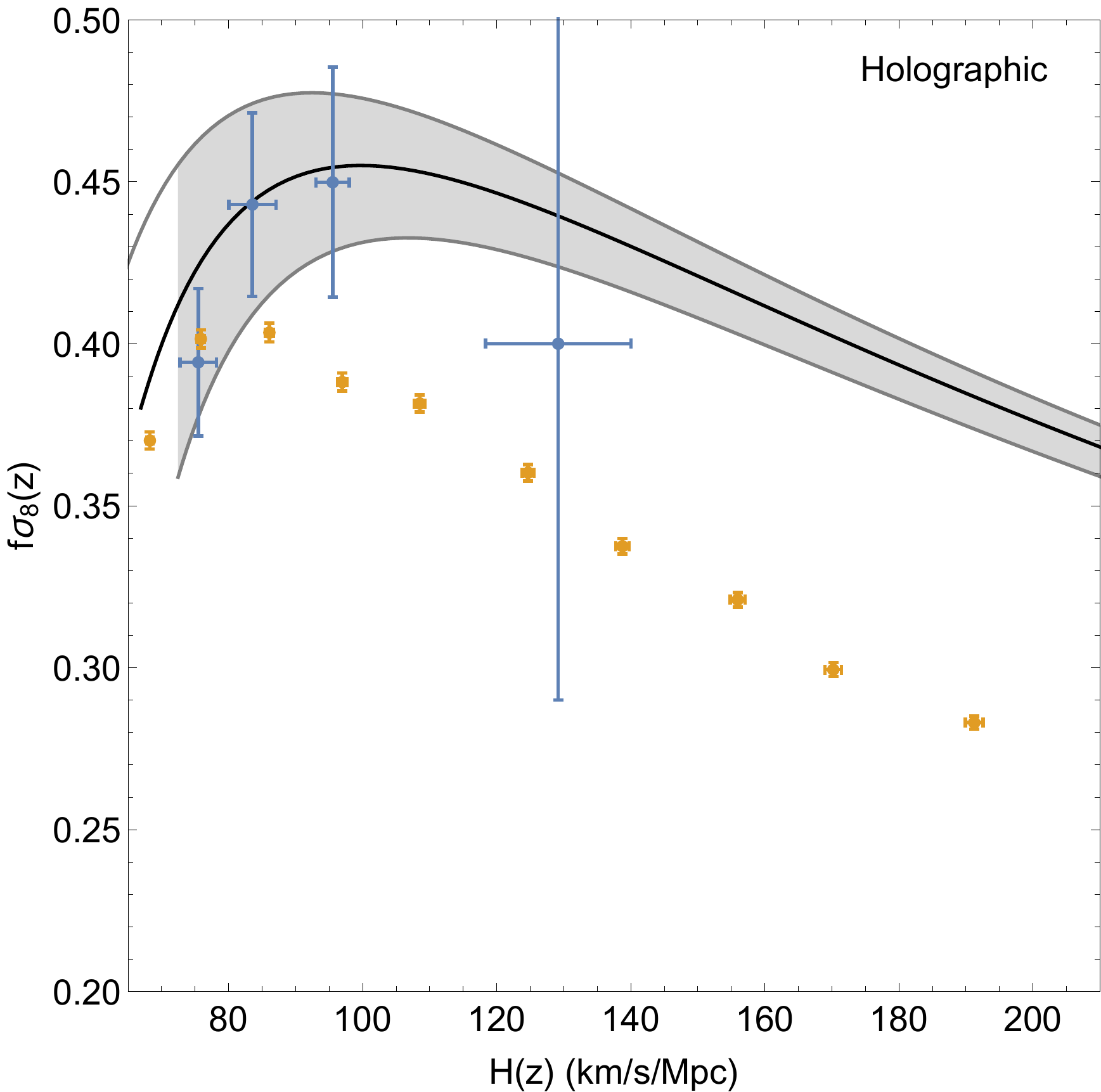}
\caption{The conjoined plots of the cosmic growth $f\sigma_8$ vs the cosmic expansion history $H(z)$ based on the real data (blue points) and mock data (yellow points), as described in the text, for the $\Lambda$CDM, $w$CDM, $w_0 w_a$CDM and Holographic models. We also show the 1$\sigma$ error regions in grey.}
\label{fig:plots1}
\end{figure*}

\begin{figure*}[!t]
\centering
\includegraphics[width = 0.45\textwidth]{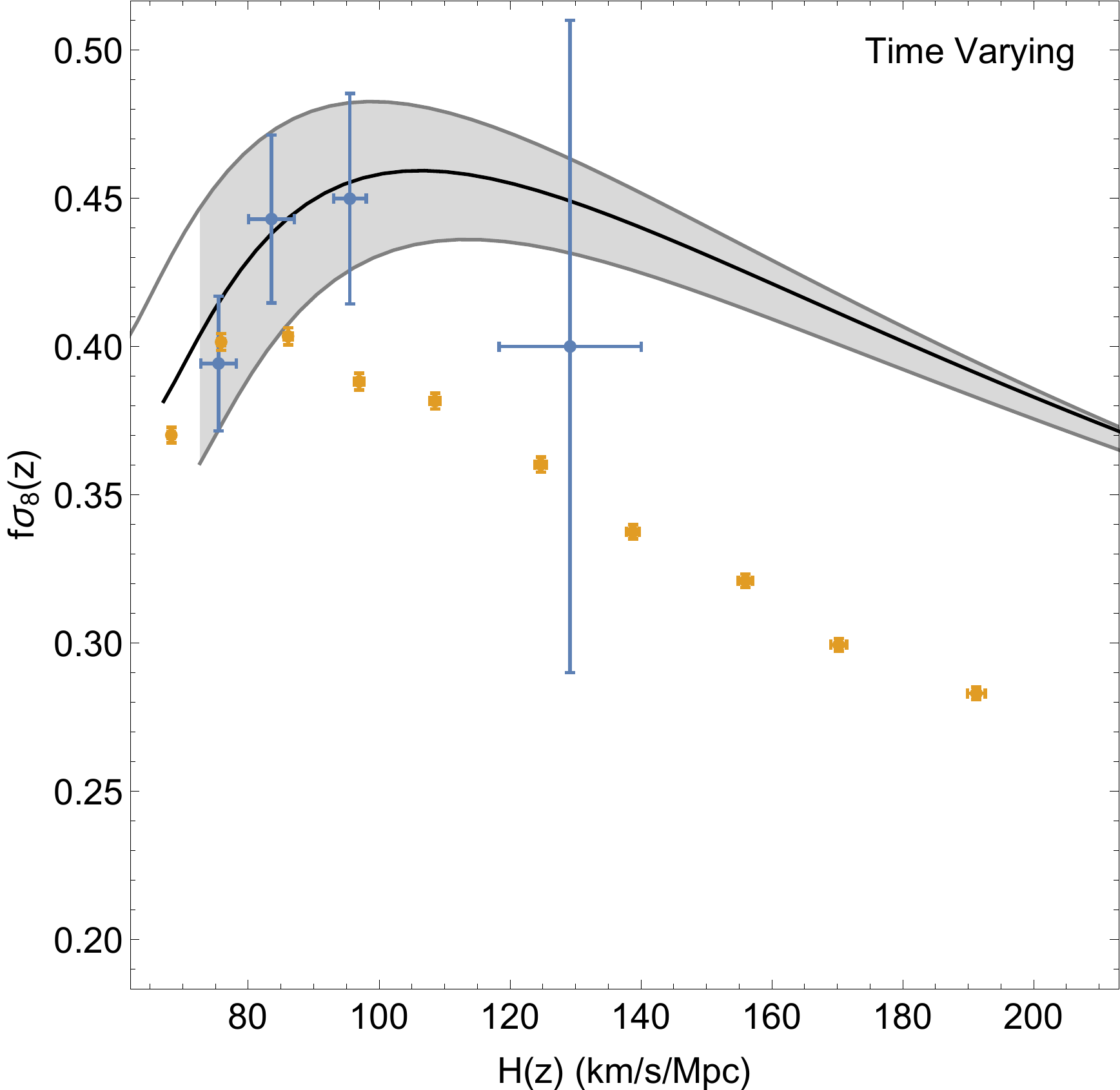}
\includegraphics[width = 0.45\textwidth]{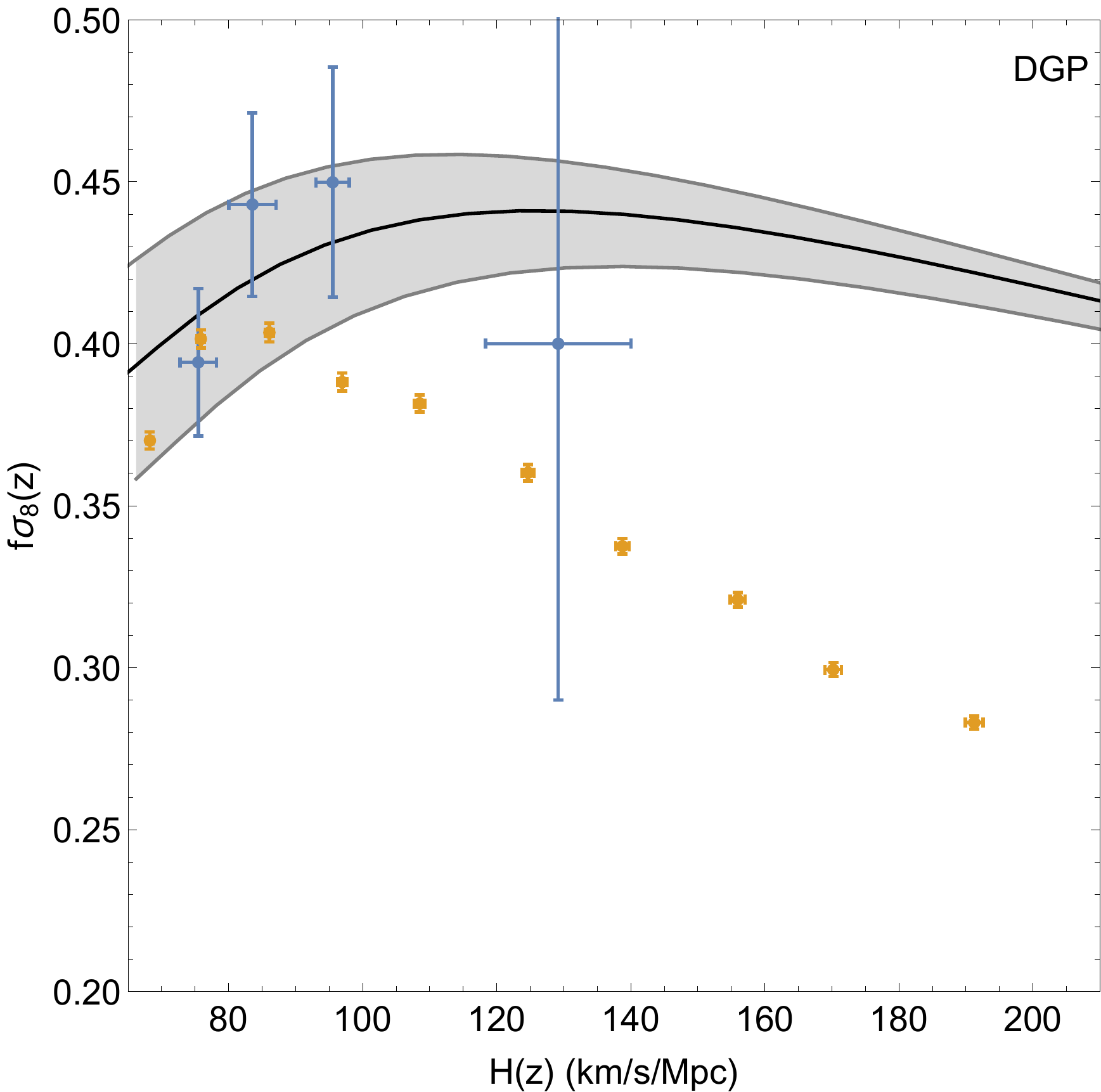}
\includegraphics[width = 0.45\textwidth]{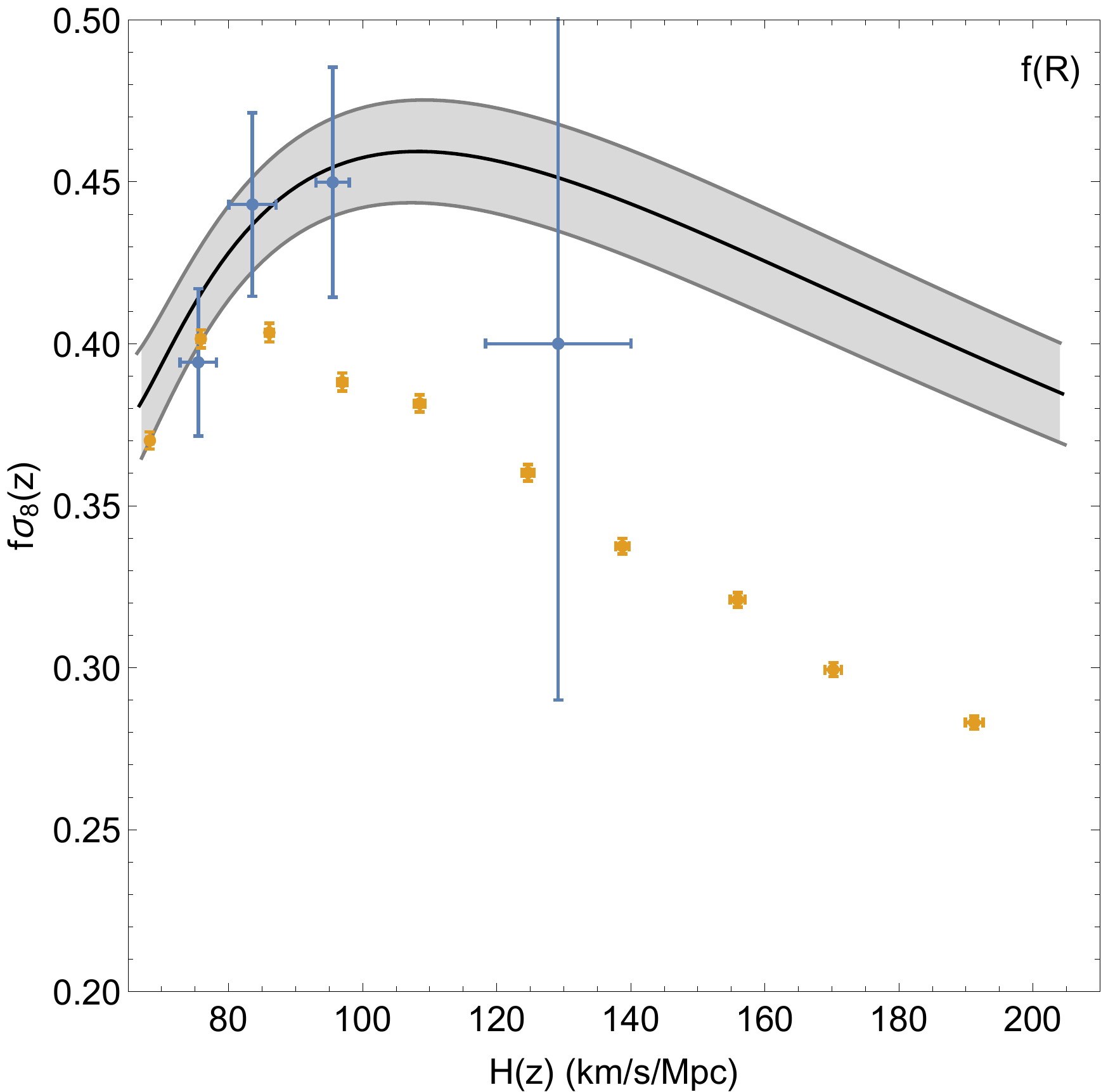}
\includegraphics[width = 0.45\textwidth]{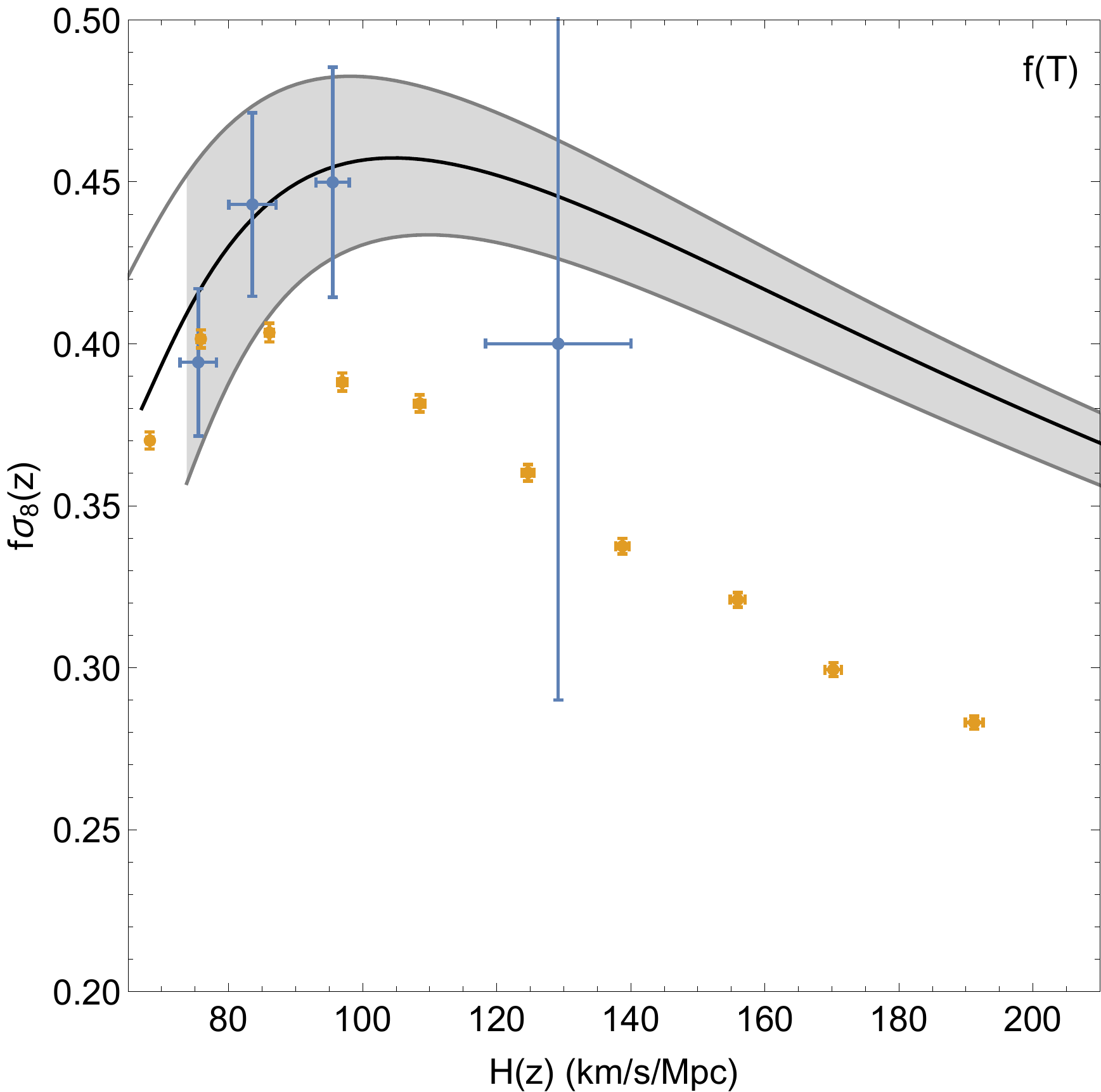}
\caption{The conjoined plots of the cosmic growth $f\sigma_8$ vs the cosmic expansion history $H(z)$ based on the real data (blue points) and mock data (yellow points), as described in the text, for the Time Varying Vacuum, DGP, $f(R)$ and $f(T)$ models. We also show the 1$\sigma$ error regions in grey.}
\label{fig:plots2}
\end{figure*}

Therefore, assuming that the Universe is isotropic, our aim is to correlate the growth as measured in the form of $f\sigma_{8}(z)$ with that of the cosmic expansion via the Hubble measurements $H(z_{\rm h})$ from the cosmic chronometer data, which are measured in the range $0.07\leq z_{\rm h}\leq 2.34$ (see Table \ref{tab:Hzdata}). More specifically, we implement the following steps:

\begin{enumerate}
\item We note that the maximum redshift for the $f\sigma_{8}$ data is $z_{max}=1.4$; hence, we only keep the $H(z)$ data up to that redshift, so that we have a common sample in the same redshift range.
\item We then split both the $f\sigma_{8}$ and the remaining $H(z)$ data in four equal bins\footnote{We found by trial and error that this is the optimal number of bins given the number of data in both sets.} and weight the data according to their errors, see Ref.~\cite{Nesseris:2014vra}.
\item For the combined sample we perform a statistical analysis and by using the models and best-fit parameters as mentioned in the previous section, we make the conjoined $H-f\sigma_{8}$ curves as seen in Figs.~\ref{fig:plots1} and \ref{fig:plots2}, along with real and mock data.
\item For the region defined by the $1\sigma$ errors of the best-fit theoretical models in the $H(z)-f\sigma_{8}(z)$ space, we also calculate the Figure of Merit (FoM), which is equal to the inverse of the enclosed area up to the maximum redshift range covered by both models. Then we rank the models according to their constraining power (see Table \ref{tab:FoM}). Note that the higher the FoM, the more constraining the model becomes.
\end{enumerate}

\begin{table}[!t]
%\centering
\caption{The FoM in the $H(z)-\fs(z)$ parameter space in descending order from higher FoM (more constraining) to lower FoM (less constraining).\label{tab:FoM}}
\begin{tabular}{cc}
%\hline
%\hline
Model & FoM \\
\hline
$\Lambda$CDM    \hspace{5pt}&\hspace{5pt} $0.3767$ \\
$f_R$CDM        \hspace{5pt}&\hspace{5pt} $0.3733$ \\
$w$CDM          \hspace{5pt}&\hspace{5pt} $0.3589$ \\
CPL             \hspace{5pt}&\hspace{5pt} $0.3488$ \\
HDE             \hspace{5pt}&\hspace{5pt} $0.2509$ \\
$\Lambda_t$CDM  \hspace{5pt}&\hspace{5pt} $0.2416$ \\
DGP             \hspace{5pt}&\hspace{5pt} $0.2288$ \\
FRDE            \hspace{5pt}&\hspace{5pt} $0.2288$ \\
$f_T$CDM        \hspace{5pt}&\hspace{5pt} $0.2223$
%\hline
%\hline
\end{tabular}
\end{table}

Following the previous steps, we now show in Figs. \ref{fig:plots1} and \ref{fig:plots2} the conjoined plots of the cosmic growth $f\sigma_8(z)$ vs the cosmic expansion history $H(z)$ based on the real data (blue points) and mock data (yellow points), for the $\Lambda$CDM, $w$CDM, $w_0 w_a$CDM, Holographic, Time Varying Vacuum, DGP, $f(R)$ and $f(T)$ models. We also show the 1$\sigma$ error regions in grey and we use them later on to calculate the $H(z)-\fs(z)$ FoM. In these plots, the real data of Table \ref{tab:fsigma8data} were split into four equally spaced bins, given in Table \ref{tab:binned}, while the mock data (yellow points) were based on an LSST-like survey with data in the range $z\in[0,2]$ and a Planck 15 best fit cosmology with $\Omega_m$=0.317, $H_0=67.2 ~ \textrm{km/s/Mpc}$, $w=-1$ and $\sigma_8=0.687$. Finally, the mock data were slit into ten equally spaced equally spaced redshift bins.

As can be seen, there is little difference between the theoretical curves of the various models, which is in agreement with our results in Ref.~\cite{Basilakos:2016nyg}. One exception however, is the DGP model which seems to be in strong tension with the data having a $\delta \chi^2\sim 100$ with the rest of the models, again confirming our earlier analysis. However, an LSST-like survey, as seen by the mock data (yellow points in Figs. \ref{fig:plots1} and \ref{fig:plots2}) will be able to discriminate the models and provide stringent constraints on the cosmic growth and expansion history. Furthermore, a survey like this will be able to shed light on the existing tension with Planck by providing data points with very small errors and by minimizing the systematics that occur by combining data from many surveys.

We also calculate the FoM in the $H(z)-\fs(z)$ parameter space, in order to assess how constraining the joined analysis of these data can be. We have defined the FoM as the inverse of the enclosed area of the 1$\sigma$ error regions in Figs. \ref{fig:plots1} and \ref{fig:plots2} up to the common maximum redshift $z_{max}=1.4$. We then rank all the models according to their constraining power, noting that the higher the FoM, the more constraining the model becomes.

The result of this comparison is shown in Table \ref{tab:FoM}, where we notice that the $\Lambda$CDM model seems to be much more constraining, followed up closely by the $f_R$CDM and $w$CDM models. As mentioned before, the DGP/Finsler models are at the bottom of the ranking, with almost half the constraining power of the rest of the models. Finally, we note that the $f(T)$ model is also at the bottom of the list, something which allows us to discriminate it from the rest of the models even though the model itself is just a perturbation around $\Lambda$CDM. Therefore, the technique of the conjoined analysis of both the $H(z)$ and $\fs(z)$ can in principle discriminate some modifications of gravity from $\Lambda$CDM.

\section{Conclusions}
\label{conclusions}
In our present study we implemented the $H-f\sigma_{8}$ conjoined method of Ref.~\cite{Linder:2016xer} in order to test the viability of a large family of DE models, including the $\Lambda$CDM, $w$CDM, $w_0 w_a$CDM, Holographic, Time Varying Vacuum, DGP, $f(R)$ and $f(T)$ models. First, we combined the available cosmic chronometer and growth data, given in Tables \ref{tab:Hzdata} and \ref{tab:fsigma8data}, respectively, and then we identified a common subsample which extends up to redshift $z_{max}=1.4$ and we binned the corresponding data. With this subsample and using the best-fit parameters of the models, as described in Table \ref{tab:bestfit}, we then performed the conjoined analysis of the $H-f\sigma_{8}$ parameters, the results of which can be seen in Figs. \ref{fig:plots1} and \ref{fig:plots2}.

We found that even though there is little difference between the theoretical curves of most of the models, something which is in agreement with our previous results in Ref.~\cite{Basilakos:2016nyg}, we note that the DGP/Finsler class of models seems to be in somewhat strong tension with the data. Also, by analyzing all the data (CMB, SnIa, BAO, growth and $H(z)$), we also observe a strong tension between the growth data and Planck. As mentioned, this tension pushes the values of the growth index $\gamma_0$ and $\sigma_8$ to be higher than our previous separate analysis.

Furthermore, using a set of mock data, based on an LSST-like survey with Planck15 cosmological parameters, i.e. the yellow points in Figs. \ref{fig:plots1} and \ref{fig:plots2}, we determined that in the near future we will be able to discriminate the models and provide even stringent simultaneous constraints on the cosmic growth and expansion history.

Using our analysis, we could also quantify the level of the constraining power of these models by using the FoM on the conjoined parameter space of $H-f\sigma_{8}$. The results of this analysis were shown in Table~\ref{tab:FoM}. By ranking the models according to their performance, noting that the higher the FoM, the more constraining the model becomes, we found that the $\Lambda$CDM model seems to be the most constraining, followed up closely by the $f_R$CDM and $w$CDM models. Again, in agreement with our previous results, we also found that the DGP/Finsler models are at the bottom of the ranking, with almost half the constraining power of the rest of the models.

Finally, we observed that the $f(T)$ model is also at the bottom of the list, something which hints that the conjoined analysis is capable of discriminating such modified gravity models from $\Lambda$CDM, thus proving to be a useful tool given the plethora of dark energy or modified gravity models that are currently available, but also the wealth of data expected from the upcoming surveys in the next decade or so.

\section*{Acknowledgements}
The authors would like to thank E. V. Linder for useful comments and suggestions. S.~B. acknowledges support by the Research Center for Astronomy of the Academy of Athens in the context of the program ``{\it Testing general relativity on cosmological scales}'' (Ref. No. 200/872).

S.~N. acknowledges support by the Research Project of the Spanish MINECO under Grant No. FPA2013-47986-03-3P, the Centro de Excelencia Severo Ochoa Program through Grant No. SEV-2012-0249, and the Ram\'{o}n y Cajal program through Grant No. RYC-2014-15843.

\raggedleft
\bibliography{bibliography}
\end{document}